\documentclass{aa}
\usepackage{graphicx}
\usepackage{natbib}
\usepackage{txfonts}
\bibpunct{(}{)}{;}{a}{}{,}
\def\al{&\!\!\!\!}
\begin{document}

\title{Fast kink modes of longitudinally stratified coronal loops}
\author{H. Safari\inst{1} \and S. Nasiri\inst{1,2} \and Y. Sobouti\inst{1}
} \institute{Institute for Advanced Studies in Basic Sciences,
Gava Zang, P O Box 45195-1159, Zanjan, Iran \and Department of
Physics, Zanjan University, Zanjan, Iran }
 \offprints{H.
Safari,\email{ hsafary@iasbs.ac.ir}}
\date{Received / Accepted }
\abstract{}
{  We investigate the standing kink modes of a cylindrical model
of coronal loops. The density is stratified along the loop axis
and changes  discontinuously at  the surface of the cylinder. The
periods and mode profiles are studies with their deviation from
those
 of the unstratified loops. The aim is to extract information on
 the density scale heights prevailing in the solar corona. }
 { The problem is reduced to solving a single
 second-order partial differential equation for $\delta B_z(r,z)$, the longitudinal
 component of the Eulerian perturbations of the magnetic field. This equation, in turn, is separated into
 two second-order ordinary, differential equations in $r$ and $z$ that are, however, connected through a
 dispersion relation between the
 frequencies and the longitudinal wave numbers.
 In the thin tube approximation, the eigensolutions are
 obtained by a perturbation technique, where the
 perturbation parameter is the density stratification
 parameter. Otherwise the problem is solved numerically.
}
 {  1) On functional dependencies of the dispersion relation the radial wave number
 is independent of the longitudinal stratification.
 2) We verify the earlier computational finding
 that the first overtone frequencies increase with increasing stratification and the
 observational finding (from analysis of TRACE data)
  that the ratio of the first to the fundamental overtone frequency decreases with
 increasing stratification. The method we use to arrive at these conclusions, however, is
 more analytical than computational, and yet our numerical results agree with the earlier results.
 3) The mode profiles depart from the sinusoidal mode
 profiles of the unstratified loops. This departure and its
 dependence  on the scale height is obtained, and might
 serve to determine scale heights once high resolution data become
 available.  } {} 

\titlerunning{Oscillations of longitudinally stratified coronal loops}
\authorrunning{Safari et al.}
\maketitle
\section{Introduction}
  Since the earliest identification of the kink oscillations
in coronal loops by Aschwanden et al. (1999a) and Nakariakov et
al. (1999), a considerable amount of data
 has been analyzed by Aschwanden et al. (2002), Schrijver
et al. (2002),  and Wang \& Solanki (2004) using the
high-resolution observations of TRACE, SoHO, Yohkoh, etc. See
Aschwanden  (2003) and   Nakariakov \& Verwichte (2005) for an
extended review   of observations of coronal  oscillations.

  Verwichte et al. (2004) report periods,  phases,  damping
times, and  mode profiles  for nine coronal loops. As expected,
the results differ from those based on simplified theoretical
models assuming cylindrical geometries, constant cross sections,
constant magnetic fields, constant gravitation, isothermal
structures, constant densities,  no initial flows, etc.

There are numerous attempts to arrive at reasonably realistic
models where various effects have been from loop geometry to
structuring and damping have to be addressed. Only after these
studies can we conclude what may or may not be important in the
context of solar magneto-seismology, i.e. for solar coronal
oscillations.  Both Smith et al. (1997) and Van Doorsslare et al.
(2004) have studied the effect of the loop curvature on the
oscillations frequencies.   Bennett et al. (1999) and Erd\'{e}lyi
\& Fedun (2006) studied the  twisted magnetic flux tubes in
incompressible media and compare their body, surface, and hybrid
modes with those of the untwisted cases. Terra-Homem et al. (2003)
went on to give a detailed discussion of the frequency shifts
caused by field-aligned background flows. Nasiri (1992) simulated
a variable cross section by assuming a long narrow-wedge geometry.
Ruderman (2003) removed the degeneracy inherent in loops of
circular cross sections by assuming elliptical cross sections.
D\'{i}az et al. (2001) studied the fast oscillations in the fine
structure of prominence fibrils. Erd\'{e}lyi \& Carter (2006) then
obtained a full analytical dispersion relation for the propagation
of MHD waves in structured magnetic flux tubes embedded within a
straight vertical magnetic environment. Mikhalyaev \& Solovev
(2005) consider the MHD oscillations of double magnetic flux tubes
in uniform external fields. De Pontieu et al. (2003a, b) analyze
the mechanism of leakage from the photosphere and the chromosphere
into the transition regions and the corona. D\'{i}az et al. (2004)
introduce photospheric line-tying boundary conditions to emphasize
the rate of leakage in damping of the oscillations. This is  an
extension of the infinite homogenous loops of Edwin \& Roberts
(1983). Mendoza - Brice\~{n}o et al. (2004) studied the effect of
the gravitational stratification, and find
 a $10-20\%$ reduction in damping times of oscillations.

Andries et al. (2005a, b) calculate damping rates of
longitudinally stratified cylindrical loops and conclude that the
ratio of the frequency of the first overtone to that of the
fundamental mode is less than 2, the value for the unstratified
loops.  They use this ratio to estimate the amount of the
density-scale
 height in the solar atmosphere.
Dymova \& Ruderman (2005) reduce the MHD equations prevailing in a
thin and longitudinally stratified magnetic fibrils, into a
Sturm-Liouville problem for the eigenvalues and eigenmodes of the
fibril. Erd\'{e}lyi \& Verth (2007) use  the approach of  Dymova
\& Ruderman (2005) to study the deviations of the mode profiles,
i.e. the eigenfunctions, of the stratified loops from the
sinusoidal profiles.

   In this paper we study the  kink modes of a
longitudinally density-stratified loop. We reduce the MHD
equations to a wave equation with variable Alfv\'{e}n speed  for
the $z$- component of the  magnetic field. The dispersion
relation, relating the frequencies and the longitudinal wave
numbers, is similar in form to that for unstratified loops. In the
thin-tube approximation, our approach converge to that of  Dymova
\& Ruderman (2005), who have rescaled the MHD equations from the
beginning to accommodate  thin tubes.

Equation of motions, boundary conditions, and the radial solutions
are dealt with in  Sects. \ref{sec2} \& \ref{sec3}. The thin tube
approximation is treated in Sec \ref{thin}.  Concluding remarks
are given in Sect. \ref{conc}.
\section{Equations of motions}\label{sec2}
\begin{center}
\begin{figure}
\includegraphics{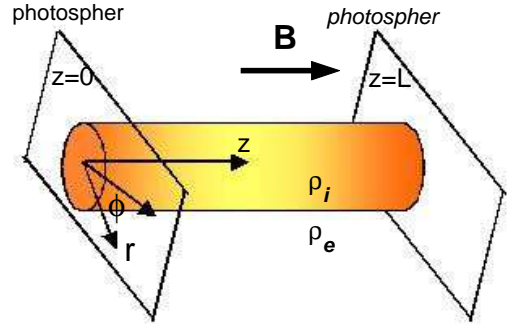}
      \vspace{4.8cm}
          \caption[]{ A sketch of the equilibrium model of the flux tube.
           Density varies along the cylinder axis and is symmetric about the midpoint.
           The magnetic field is uniform along the $z$-axis.}
\label{tube}
   \end{figure}
   \end{center}

  A coronal loop, with its ends at the photosphere and
 with a relatively small curvature ( that is, the radius of curvature of the
 loop much larger than the loop length ), is idealized as a circular cylinder.
 The cylinder is assumed to have no initial material flow,
 to be pervaded by a uniform magnetic field along its axis, $\mathbf{B}=B\hat{z}$, and to have
 negligible gas pressure ( zero-$\beta$  approximation ). The
 length and radius of the loop are  $L$ and  $a$, respectively, (see Fig. \ref{tube}).
 The coordinate system is the cylindrical one, $(r,\phi,z)$.
  The density is assumed to be
\begin{eqnarray}\begin{array}{cccc}
                  \rho(\epsilon,z) &= &\rho_i(\epsilon)f(\epsilon,z), ~~~& r< a \\
                          & = &\rho_e(\epsilon)f(\epsilon,z), ~~~& r> a \\
                \end{array} \label{rho}
\end{eqnarray}
\begin{equation}
f(\epsilon,z)=exp(-\frac{\epsilon}{\pi}\sin\frac{\pi
z}{L})\label{density},
\end{equation}
where $\rho_i(\epsilon)$ and $\rho_e(\epsilon)$ are the interior
and exterior densities at the footpoints of the loop,
$\epsilon=L/H$, where $H$ indicated the scale length ( see bellow
). The density variations for inside and outside of the loop are
governed by the same function $f(\epsilon,z)$. The exponential
stratification of the density has been pointed out by Aschwanden
et al. (1999b) on the basis of their EUV studies in 30 loops from
the SoHO/EIT data. The sinusoidal form of the exponent is
suggested by Andries et al. (2005b). It  ensures the symmetry with
respect to the midpoint of the loop. See, however, Erd\'{e}lyi \&
Verth (2007) for alternatives to this sinusoidal exponent.

Here, as in most other works, the stratification of the density
and its exponential scaling is adopted as an empirical fact.
Evidently its source is not a gravitational one, because, a) the
gravitational scale height in coronal conditions far exceeds any
length scale in solar environments and b) the loops are not,
generally, oriented along the gravitational field of the sun.

In the following we consider loops of different scale heights but
of constant total column masses  $\mu_i$ and $\mu_e$, independent
of $\epsilon$. Thus,
\begin{eqnarray}\al\al \frac{\mu_i}{\rho_i(\epsilon)}=\frac{\mu_e}{\rho_e(\epsilon)}=\int_0^Lf(\epsilon,z)dz=
L[I_0(\frac{\epsilon}{\pi})-\bf{L}_0(\frac{\epsilon}{\pi})],~~~
\end{eqnarray}
where $I_0$ is the  modified Bessel function of the first kind and
and $\bf{L}_0$ is the modified Struve function (see Gradshteyn \&
Ryzbik, 2000).
The linearized   ideal MHD equations for   the Eulerian
perturbation  in the velocity and the magnetic fields in a
zero-$\beta$ plasma are
\begin{equation}
   \frac{\partial\delta\mathbf{v}}{\partial
            t}=\frac{1}{4\pi\rho}\{(\mathbf{\nabla}\times\delta\mathbf{B})\times\mathbf{B}+(\mathbf{\nabla}\times\mathbf{B})\times\delta\mathbf{B}
            \}\label{1},
   \end{equation}
   \begin{equation}
 \frac{\partial\delta\mathbf{B}}{\partial
            t}=\mathbf{\nabla}\times(\delta\mathbf{
            v}\times\mathbf{B})\label{2}.
   \end{equation}
  Note  that Eq. (\ref{2}) ensures the
Gaussian law, $\nabla. ~\delta\mathbf{B}=0$.
  From Eqs. (\ref{1}) and (\ref{2}) one can show, after some
 straightforward calculations that
\begin{eqnarray}
\al\al \left(\frac{\omega^2}{v_A^2}+\frac{\partial^2}{\partial
z^2}-\frac{m^2}{r^2}\right)\frac{\delta
B_z}{B}=\left(\frac{\omega^2}{v_A^2}+\frac{\partial^2}{\partial
z^2}\right)\frac{1}{r}\frac{\partial}{\partial
r}\left(r\frac{\delta
v_r}{i\omega}\right),\label{dvdb2}\\
\al\al \left(\frac{\omega^2}{v_A^2}+\frac{\partial^2}{\partial
z^2}\right)\frac{\delta v_r}{i\omega}=-\frac{\partial}{\partial
r}\frac{\delta B_z}{B} \label{dvdb1},
\end{eqnarray}
where $v_A(z,\epsilon)=B/\sqrt{4\pi\rho(z,\epsilon)}$ is the local
Alfv\'{e}n speed and is different for inside and outside of the
loop. The term $\delta v_r/i\omega$ appearing in Eqs.
(\ref{dvdb1}) and (\ref{dvdb2}) is actually the   Lagrangian
displacement vector in the loop.
 The remaining components of
$\delta\mathbf{v}$ and $\delta\mathbf{B}$ are given by
\begin{eqnarray} \delta v_\phi&=&\frac{\omega}{mB_z}r\delta
B_z-\frac{1}{im}\frac{\partial}{\partial r}(r\delta v_r) ,~~~
\delta v_z=0,\\ \delta
B_r&=&-\frac{B_z}{i\omega}\frac{\partial\delta v_r}{\partial z},
~~~~\delta B_\phi=-\frac{B_z}{i\omega}\frac{\partial\delta
v_\phi}{\partial z}.
\end{eqnarray}
See, e. g.,  Karami et al (2002) and Safari et al. (2006) for
details. Let us also emphasize that the present form of Eq.
(\ref{dvdb1}) utilizes the fact that $\rho$ and, consequently,
$v_A$ depend only on $z$. There is no radial  variation except for
a step discontinuity at the surface of the tube.

Eliminating $\delta v_r$ between Eqs. (\ref{dvdb1}) and
\ref{dvdb2}) gives
\begin{eqnarray} \left(\frac{1}{r}\frac{\partial}{\partial
r}r\frac{\partial}{\partial r}+\frac{\partial^2}{\partial
z^2}-\frac{m^2}{r^2}\right)\delta B_z+\frac{\omega^2}{v_A^2}\delta
 B_z=0\label{dbz1}.
\end{eqnarray}
 This is a wave equation for $\delta
B_z$ with the variable speed $v_A(\epsilon,z)$. We solve it by the
separation of variables.  Let $ \delta B_z=R(r)Z(z)$.  For later
convenience, we add and subtract the term
$\frac{\omega^2}{v_A^2(\epsilon=0)}\delta B_z$ to and from Eq.
(\ref{dbz1}), and follow the usual procedure for the separation of
variables. We obtain
\begin{eqnarray}
\al\al \frac{1}{R}\frac{1}{r}\frac{d^2R}{dr^2}+\frac{1}{R}\frac{d R}{dr}-\frac{m^2}{r^2}+\frac{\omega^2}{v_A^2(\epsilon=0)}\nonumber\\
\al\al
~~=-\frac{1}{Z}\frac{d^2Z}{dz^2}-\omega^2\left(\frac{1}{v_A^2(\epsilon,z)}-\frac{1}{v_A^2(\epsilon=0)}\right)
=\kappa_z^2,\label{dbzz}
\end{eqnarray}
  where $\kappa_z^2(\epsilon)$ is the constant of separation; and
in the absence of longitudinal stratification, it reduces to the
longitudinal wave number. Equation  (\ref{dbzz}) may now be
written as
\begin{eqnarray}
\al\al
\left(\frac{d^2}{dr^2}+\frac{1}{r}\frac{d}{dr}-\frac{m^2}{r^2}
\right)R(r)+k^2R(r)=0,\label{psi1}\\
\al\al \left(\frac{d^2}{dz^2}-k^2\right)Z(z)+
\frac{\omega^2}{v_A^2}Z(z)=0,~~k^2=\frac{\omega^2}{v_A^2|_{\epsilon=0}}-\kappa_z^2.\label{phi1}
\end{eqnarray}
  Both Eqs.
(\ref{psi1}) and (\ref{phi1}) are actually a pair for inside,
$r<a$, and outside, $r>a$, of the tube. They are to be solved
simultaneously for $R$, $Z$, $\kappa_z$, and $\omega$.

\subsection{Boundary conditions}\label{bn}
\begin{enumerate}
 \item  The  changes in
    total pressure
should be continuous. On account of the zero-$\beta$ approximation
and constancy of $B$, this reduces to the requirement of the
continuity of $\delta B_{z}$. Thus
\begin{eqnarray}
\al\al R^{\rm{interior}}(a)=R^{\rm{exterior}}(a),\label{boundary1a}\\
\al\al Z^{\rm{interior}}=Z^{\rm{exterior}},~~~~{\rm for~ all~
}z.\label{boundary1b}
\end{eqnarray}
    \item  On account of
$\mathbf{\nabla}\cdot\delta\mathbf{B}=0$, $\delta B_{r}$ should be
continuous at $r=a$.  This, gives  (Karami et al 2002)
\begin{equation}
\left.\frac{1}{k_{i}^2}\frac{dR^{\rm{interior}}(k_{i}r)}{dr}\right|_{r=a}=-\left.
\frac{1}{k_{e}^2}\frac{dR^{\rm{exterior}}(k_{e}r)}{dr}\right|_{r=a}.
\label{boundary2}\end{equation}
\item  The footpoints, $z = 0 ~\&~L$, are expected to be nodes. This imposes the
conditions
\begin{equation}
Z(z=0~\&~L)=0\label{boundaryz}.
\end{equation}
Equations (\ref{boundary1a}), (\ref{boundary2}), and
(\ref{boundaryz}) give four boundary conditions for the two
second-order differential Eqs. (\ref{psi1}) \& (\ref{phi1}).
\end{enumerate}
\section{Solutions of Eqs. (\ref{psi1}) and (\ref{phi1})}\label{sec3}
 Interior solutions of Eq.
(\ref{psi1})  that are regular at $r=0$ are $J_m(|k_i|r)$ for
$k_i^2>0$ or $I_m(|k_i|r)$ for $k_i^2<0$. Exterior solutions that
decay with $r\to\infty$ are $K_m(|k_e|r)$. They occur for
$k_e^2<0$ and are evanescent waves.

 Imposing the boundary conditions of Eqs.
(\ref{boundary1a}) and (\ref{boundary2}) gives
\begin{equation}
\frac{1}{k_i}\frac{J_m'(|k_i|a)}{J_m(|k_i|a)}-\frac{1}{k_e}\frac{K_m'(|k_e|a)}{K_m(|k_e|a)}=0,
\label{jk}\end{equation} where $'$ indicates a derivative of a
function with respect to its argument. The same relation holds for
surface waves with $J_m(|k_i|r)$ replaced by $I_m(|k_i|r)$. For
unstratified thin and thick tubes Eq. (\ref{jk}) is analyzed  by
 Edwin \& Roberts (1983). Here we study Eq. (\ref{jk}) for
thin stratified loops by perturbational and numerical techniques.
\section{ Thin tube
approximation}\label{thin} For $a/L\ll1$ and $m\geq1$, the
dispersion relation of Eq. (\ref{jk}) gives  $|k_i|\approx |k_e|$.
From the definition of $k^2$ in Eq. (\ref{psi1}), one then obtains
\begin{eqnarray} \omega=\kappa_z B(4\pi\bar{\rho}_0)^{-1/2}, ~\bar{\rho}_0=\frac{1}{2}[\rho_i(0)+\rho_e(0)].\label{omega1} \end{eqnarray}
This is the kink oscillation frequency in the presence of
stratification. Edwin \& Roberts (1978), Karami et al. (2002), Van
Doorsselaere et al. (2004),  and D\'{i}az et al. (2004) all
obtained a similar result for $\omega$ in unstratified flux tubes.
Here, however, $\kappa_z$ is given by Eq. (\ref{dbzz}). It reduces
to the longitudinal wave number in the absence of stratification.
Substituting Eq. (\ref{omega1}) into Eq.(\ref{phi1}) (interior
with $k_i^2>0$ and exterior with $k_e^2<0$) and using the boundary
condition of Eq. (\ref{boundary1b}) yield
\begin{eqnarray}
\al\al \frac{d^2Z}{dz^2}+\frac{4\pi\omega^2}{B^2}F(\epsilon,z)Z(z)=0,~~Z=Z^{\rm interior}=Z^{exterior},\label{phiper}\\
\al\al F(\epsilon,z)=\frac{\rho_i(\epsilon)+\rho_e(\epsilon)}{2}f(\epsilon,z).\nonumber
\end{eqnarray}
Equation (\ref{phiper}) is an eigenvalue problem weighted by
 $F(\epsilon,z)$.  It is the same as that of Dymova \& Ruderman (2005) derived,
  however, with a different approach.

  From Eq. (\ref{phiper}) one may write down the following
integral expression for $\omega^2$
\begin{equation}
\omega^2=\frac{B^2}{4\pi}\frac{\int |dZ/dz|^2dz}{\int
F(\epsilon,z)|Z(z)|^2dz}.\label{omega2}
\end{equation}
Some general properties of $\omega^2$ can be inferred from Eq.
(\ref{omega2}). In Fig. \ref{fz}, $F(\epsilon,z)$ is plotted
versus $z$ for three values of $\epsilon$. It has maxima at
footpoints and a minimum at the apex.  The larger $\epsilon$, the
higher the maxima and the lower apex become.  This reduces the
integral in the denominator of Eq. (\ref{omega2}) and causes an
increase in $\omega^2$. There is even the possibility of the
integral, and thereby $\omega^2$, becoming infinity.

In the following section, Eq. (\ref{phiper}) is solved for small
amount of the density scale heights by perturbation, and for
arbitrary scale height parameters numerically.
\subsection{Perturbation method}\label{perturbation} 
\begin{figure} 
\unitlength1cm
\begin{center}
\includegraphics[width=0.45\textwidth,clip=]{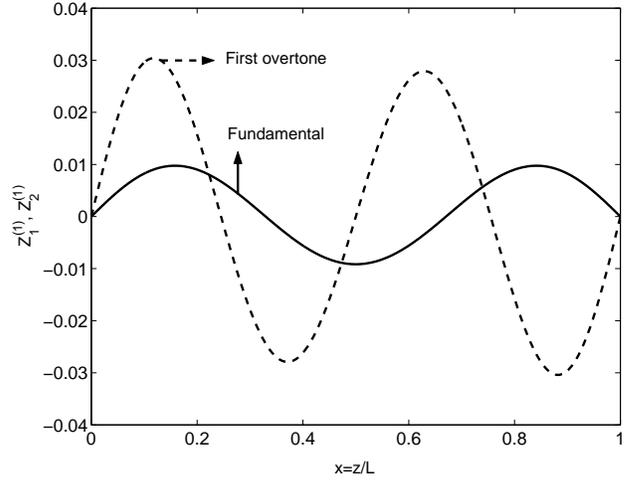}
\end{center}
\caption[1] {Plot of the first-order perturbations in amplitude
profiles of the fundamental and the first overtone modes. See Eqs.
(\ref{zerovec}) and (\ref{firstvec}). } \label{deltaz}
\end{figure}

The scale height parameter
is chosen as the perturbation parameter, and all variables and
equations are expanded in powers of $\epsilon$. Thus,
\begin{eqnarray}
\al\al \omega=\omega^{(0)}+\epsilon\omega^{(1)}+\cdots,\label{perturbed1}\\
\al\al Z(z)=Z^{(0)}(z)+\epsilon Z^{(1)}(z)+\cdots,\\
\al\al F(\epsilon,z)=\bar{\rho}_0\left[1
+\epsilon\left(\frac{2}{\pi^2}- \frac{1}{\pi}\sin\pi\frac{z}{L}
\right)+\cdots\right].\label{perturbed3}
\end{eqnarray}
 Equation (\ref{phiper}) splits into zeroth and first-order
components
\begin{eqnarray}
\al\al \frac{d^2Z^{(0)}}{dz^2}+\frac{4\pi\omega^{(0)^2}\bar{\rho}_0}{B^2}Z^{(0)}=0\label{z-order},\\
\al\al
\frac{d^2Z^{(1)}}{dz^2}+\frac{4\pi\omega^{(0)^2}\bar{\rho}_0}{B^2}Z^{(1)}
+\frac{8\pi\omega^{(0)}\omega^{(1)}\bar{\rho}_0}{B^2}
Z^{(0)}\nonumber\\
\al\al=-\frac{4\pi\omega^{(0)^2}\bar{\rho}_0}{B^2}\left(\frac{2}{\pi^2}-
\frac{1}{\pi}\sin\frac{\pi
z}{L}\right)Z^{(0)}.\nonumber\\\label{f-order}
 \end{eqnarray}
 Solutions of  Eq. (\ref{z-order}) for $Z^{(0)}$ and $\omega^{(0)}$
 with boundary conditions of Eq. (\ref{boundaryz}) are
\begin{eqnarray}
\al\al  \omega^{(0)}_l=\frac{l\pi}{L}B~(4\pi\bar{\rho}_0)^{-1/2}~~l=1,2,3,...,\label{omega0}\\
\al\al  Z_l^{(0)}(z)=\sqrt{\frac{2}{L}}\sin \frac{l\pi}{L}
z,\label{zerovec}
\end{eqnarray}
where $l$ is the longitudinal mode number,
 and $\omega^{(0)}_l$  the kink mode frequency in the absence
of stratification. The right hand side of Eq. (\ref{f-order}) is
now known. Multiplying it by $Z^{(0)*}$, integrating over $z$, and
reducing it by Eq. (\ref{z-order}) gives the first-order
corrections
\begin{eqnarray}
\al\al {\omega}^{(1)}_l=\omega^{(0)}_l\frac{1}{2}(I_{ll}-\frac{2}{\pi^2})\label{o1}\\
\al\al  Z_l^{(1)}=\sum_{l'\neq
l}c_{ll'}Z^{(0)}_{l'},~~c_{ll'}=\frac{l^2I_{ll'}}{l^2-l'^2},
\label{firstvec}\end{eqnarray}
 where
\begin{eqnarray}
I_{ll'}&=&\int_0^LZ^{(0)*}_{l'}\frac{1}{\pi}sin\frac{\pi
z}{L} Z^{(0)}_ldz\nonumber\\
&=&-\frac{4}{\pi^2}\frac{ll'(1+\cos l\pi \cos
l'\pi)}{l^4+(-1+l'^2)^2-2l^2(1+l'^2)} \label{integ}.
\end{eqnarray}
Equation (\ref{integ})  agrees with the result of Andries et al.
(2005a) ( see $S_n$ in their Eqs. (3) and (4)). The Sheffield
school maintains that a knowledge of the deviations of the
amplitude profile of stratified loops from those of the
unstratified one, $Z_l-Z_l^{(0)}\approx\epsilon Z_l^{(1)}$, in our
notation can give information on the density stratification of the
loop (private communication, see also Erd\'{e}lyi \&  Verth 2007).
In Fig. \ref{deltaz}, $Z^{(1)}_1$ and $Z^{(1)}_2$ are plotted as
functions of $z$. The first, $Z^{(1)}_1$, exhibits two maxima at
$z/L=1/6~ \&~5/6$ and one minimum at $1/2$. It is zero at
$0,~1/3,~2/3,~$ and $1$. The second, $Z^{(1)}_2$, shows two maxima
at $z/L=1/8~$ and $5/8$, two minima at $3/8$ and $7/8$, and is
zero at $0,~1/4,~2/4,~3/4,$ and $1$. The maxima of
$Z_1^{(1)}/Z_1^{(0)}$ and $Z_2^{(1)}/Z_2^{(0)}$ are 0.02 and 0.04,
respectively; see Table \ref{tab1} column 2 \& 4. From the TRACE
data, Aschwanden et al. (2002) report a displacement amplitude of
100-8800 km at the apex of the coronal loops. For a loop of
$L=$100 Mm, $\epsilon=L/H=2 $, the calculated percentages,
0.02-0.04, ${\rm {Max}}(\Delta Z_1)$ fall in the range 2-176 km.
One should be aware of whether the accuracies of observed data
allows the detection of such minute effects.

 We note
that $\omega^{(1)}$ is positive and tends to zero for $l\gg1$.
 The ratio of the periods of the fundamental and the first
overtone is
\begin{eqnarray}
\frac{P_1}{P_2}&=&\frac{\omega_2}{\omega_1}\nonumber\\
&=&2\frac{1+\epsilon\omega_2^{(1)}/\omega_2^{(0)}}{1+\epsilon\omega_1^{(1)}/\omega_1^{(0)}}
=2\frac{1+\epsilon\frac{1}{15\pi^2}}{1+\epsilon\frac{1}{3\pi^2}}<2.
\label{ratiothin}\end{eqnarray}  Equation (\ref{ratiothin}) is  a
useful tool to estimate the density scale height of the loops, see
also Roberts (2005).
\subsection{Numerical method}\label{numericalthin}
\begin{figure} 
\unitlength1cm
\begin{center}
\includegraphics[width=0.45\textwidth,clip=]{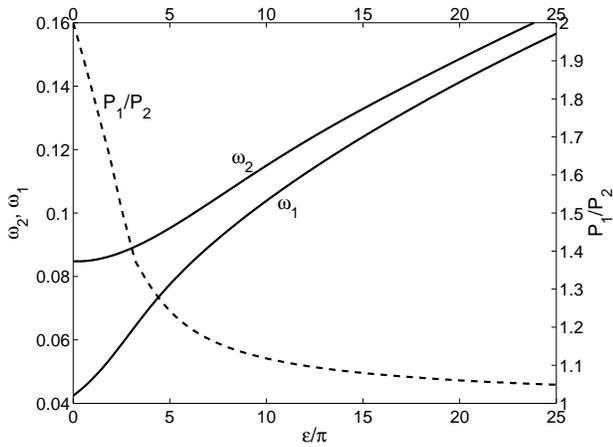}
\end{center}
\caption[1] { The frequencies, $\omega_1$ and $\omega_2$, and the
ratio $P_1/P_2=\omega_2/\omega_1$ versus $\epsilon/\pi$. Auxiliary
parameters are the tube length =$100a$, B = 100 G, and
$\rho_e/\rho_i =0.1$. }\label{p1p2}
\end{figure}

  Using a numerical code based on shooting method, Eq.
(\ref{phiper}) is solved for eigenvalues and eigenfunctions.  For
the unstratified loops, where $\rho_i$ and $\rho_e$ are constants,
the eigenfrequencies and the eigenfunctions are those of Eqs.
(\ref{omega0}) and (\ref{zerovec}), respectively. For a range of
$0<\epsilon/\pi<25$, we have calculated the fundamental and the
first overtone kink frequencies $\omega_1$ and $\omega_2$,
respectively, and the ratio $\omega_2/\omega_1=P_1/P_2$. The
results are plotted in Fig. \ref{p1p2}.  As anticipated from Eq.
(\ref{omega2}) and the behavior of $F(\epsilon,z)$, both
frequencies show monotonous increase with increasing $\epsilon$.
For small $\epsilon$, $\omega_1$ has a steeper slope than
$\omega_2$, but both approach each other as $\epsilon$ increases.
The ratio $P_1/P_2$ begins at 2 for unstratified loops,
$\epsilon=0$, and decreases to \emph{one} at large $\epsilon$.
From the TRACE data, Verwichte et al. (2004) find the ratio 1.64
and 1.81 for two of their observed loops. From Fig. \ref{p1p2},
$\epsilon$ corresponding to these ratio are 1.93$\pi$ and
1.07$\pi$, respectively. Assuming typical loop lengths,
$L=100-250$ Mm, the density scale height falls in the range of
$H\approx$16-41 and 30-74 Mm, respectively. These scale heights
agree with the finding of Andries et al. (2005a, b).

\begin{figure} 
\unitlength1cm
\begin{center}
\includegraphics[width=0.45\textwidth,clip=]{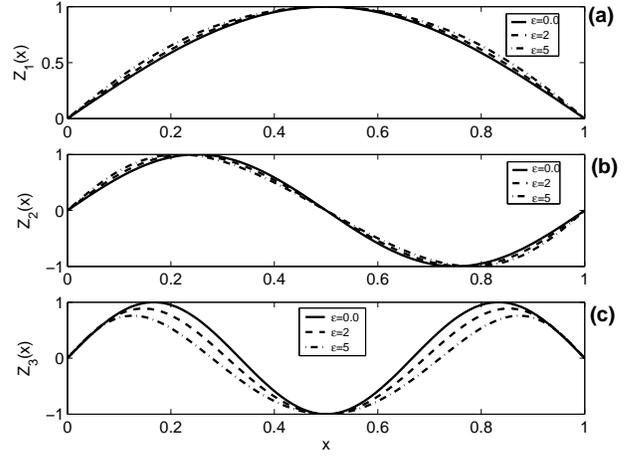}
\end{center}
\caption[1] {a) The fundamental mode ($l=1$), b) the first
overtone mode ($l=2$), and c) the second overtone modes ($l=3$)
 versus z, for unstratified, $\epsilon=0$, (
corresponding to $H=\infty$, $L$=100Mm , say ) and stratified,
$\epsilon=2,~5$, {( corresponding to $H=50$ \& 20Mm and $L=100$Mm,
say )}.} \label{phi-fig}
\end{figure}

  The longitudinal part of the eigenfield, $Z(z)$, is plotted in
Figs. \ref{phi-fig} for $l=$1, 2, 3 and $\epsilon=$0, 2, 5. As
$\epsilon$ increases, a) the eigenprofiles depart further from the
sinusoidal profiles of the unstratified case, b) the antinodes
move towards the footpoints, and c)  the central antinode gets
flattened in the case of odd $l$.

\begin{figure} 
\unitlength1cm
\begin{center}
\includegraphics[width=0.45\textwidth,clip=]{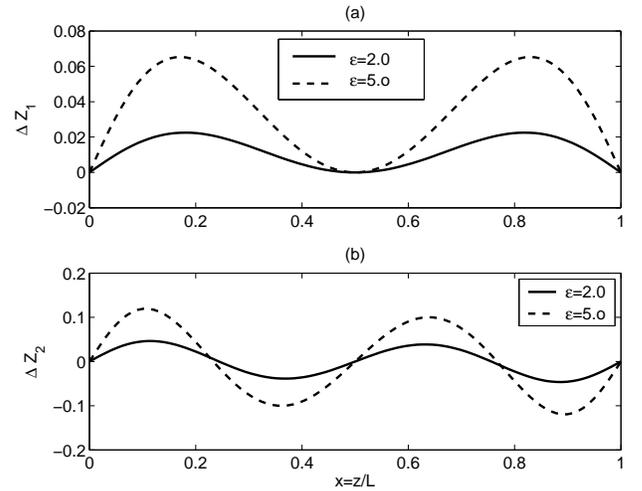}
\end{center}
\caption[1] { Amplitude differences of the fundamental and first
overtone \textbf{modes} from those of the unstratified loops for
$\epsilon=2$ \& 5, numerical calculations. } \label{deltaz_num}
\end{figure}

The differences between the eigenprofiles of the stratified and
the unstratified cases,  $\Delta
Z_l=Z_l(\epsilon,z)-Z_l(\epsilon=0,z)$, are plotted in  Fig.
\ref{deltaz_num} for  the fundamental and the first overtone
modes. Expectedly, the difference increases with increasing
$\epsilon$. The maxima rise and move towards the footpoints as
$\epsilon$ increases. For example, for $\epsilon=2$ and $5$ (
corresponding to $L=200$Mm, $H=50$ and $20$Mm, say ), the first
maximum of $\Delta Z_1$ is located at $z=$36Mm and 35Mm,
respectively,  in agreement with Erd\'{e}lyi \& Verth (2007).

Table \ref{tab1} shows  ${\rm Max}(\Delta Z_l/\epsilon Z_l)$ for
$l=$1, 2 and $\epsilon=2$, 5. Columns 2 and 4 are from the first
order perturbation calculations. Column 3 and 5 are from the full
numerical analysis.  The proximity of the two different methods of
calculations, even at scale heights as large as $\epsilon=5$, is
striking.
\begin{table}
\caption{  Maximum amplitude differences between stratified and
unstratified loops, 1st-order perturbation ( columns 2 \& 4 ), and
full numerical calculations ( columns 3 \& 5 ). } \label{tab1}
\begin{center}
\begin{tabular}{ccccc} \hline
$\epsilon=\frac{L}{H}$ &${\rm Max}\left(\frac{\Delta Z_1}{\epsilon
Z^{(0)}_1}\right)$&
${\rm Max}\left(\frac{\Delta Z_2}{\epsilon Z^{(0)}_2}\right)$ \\
&Perturbation~~~Numerical&Perturbation~~~Numerical\\
\hline \hline 2 & 0.02~~~~~~~~~~~~~0.017 &0.04~~~~~~~~~~~~~0.036   \\
  5 & 0.02~~~~~~~~~~~~0.022 &0.04~~~~~~~~~~~~~~0.04~~\\
 \hline
\end{tabular}
\end{center}
\end{table}
Aschwanden et al. (2002) report a displacement amplitude of
100-8800 km at the apex of the kink modes. Combined with the
fractional deviations of Table \ref{tab1}, one may calculate
actual physical deviations of 2-176 km for $\epsilon=2$ and
$6-528$km for $\epsilon=5$.  This result also  agrees with
Erd\'{e}lyi \& Verth (2007). The question remains as to whether
the accuracy of the observed data will allow the detection of such
small effects.

\begin{figure} 
\unitlength1cm
\begin{center}
\includegraphics[width=0.45\textwidth,clip=]{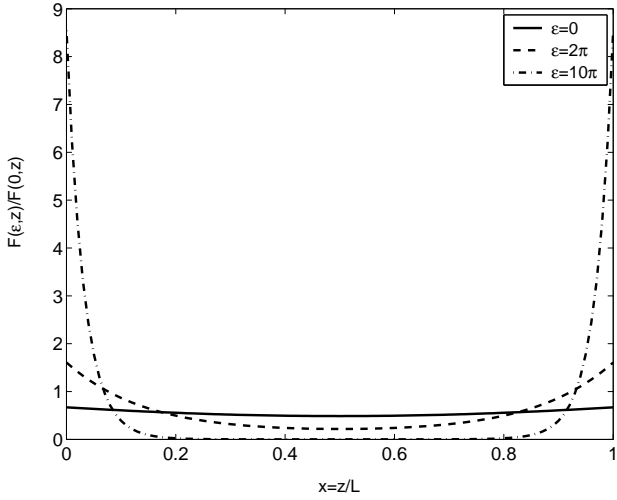}
\end{center}
\caption[1] {  $F(\epsilon,z)$ versus $z/L$.  $\epsilon=0.0$
(solid line), $\epsilon=2\pi$ (dashed line), and $\epsilon=10\pi$
(dot dashed line) . As $\epsilon$ increases, $F(\epsilon,z)$
recedes towards the end points and is void around where the mid
point widens.} \label{fz}
\end{figure}

\section{CONCLUSIONS}\label{conc}
{  We have studied the MHD oscillations of a vertically stratified
coronal loop
\begin{itemize}
\item  Equation (\ref{phiper}), for the longitudinal component of the
waves, is the same as those of Dymova \& Ruderman (2005).
\item The oscillations frequencies, obtained from Eq.
(\ref{phiper}), depend on the stratification parameter $\epsilon$.
This in turn, makes the radial wave numbers, $k_i$ \& $k_e$ of
Eqs. (\ref{phi1}) and (\ref{jk}) $\epsilon$- dependent.
\item In the thin tube approximation, the eigenfrequencies are obtained
by both perturbational  and numerical techniques.
\item  The effect of stratification is best understood by the behavior
    of $F(\epsilon, z)$, highlighted in Fig. \ref{fz}. $F(\epsilon, z)$
     is all positive. But at large $\epsilon$ it
    becomes insignificant in  broad neighborhood of $z=L/2$.
     This reduces the denominator
    in Eq. (\ref{omega2}) and lets $\omega^2$ grow. This in turn
    results  in washing-out finite time  measurements of the phenomena under study.
     \item The ratio of the periods of the fundamental and the first overtone modes (2 for
     unstratified loops) decreases markedly and approaches 1 with an increasing density-scale height parameter.
     For  $\rho_e/\rho_i=0.1$,  $\epsilon/\pi=L/\pi H=1.07$, \& $1.94$,  the ratio $P_1/P_2$ is
     $1.81$ \& $1.64$. These are in
    good agreement with the observational data of Verwichte et al.
    (2004), $1.81 \pm 0.25$ and $1.64 \pm 0.23$. The  latter are
    deductions  from TRACE observations, assuming the same density
    contrast and scale height parameter.
    \item The eigenfunctions of stratified loops deviate from
    the sinusoidal profiles of the unstratified ones. Relative deviations
    grow with $\epsilon$ and are of are close to a few percent, in general
    (see Table \ref{tab1}).

\end{itemize}
\begin{acknowledgements}
 This work was supported by
the Institute for Advanced Studies in Basic Sciences (IASBS),
Zanjan.  Authors wish to thank Profs. Robert Erd\'{e}lyi and Bernd
Inhester for their valuable consultations  and the anonymous
referee of A\&A, who meticulous suggestion have enhanced the
clarity of the paper.
\end{acknowledgements}

\end{document}